\documentclass[prl,aps,twocolumn]{revtex4}

\usepackage{psfrag}
\usepackage{graphicx}
\usepackage{dcolumn}
\usepackage{latexsym,amsfonts}
\usepackage{bm}
\usepackage{amssymb}
\begin{document}

\title{First--Forbidden Continuum- and Bound-State $\beta^-$--Decay
  Rates of Bare ${^{205}}{\rm Hg}^{80+}$ and ${^{207}}{\rm Tl}^{81+}$
  Ions}

\author{M. Faber$^{a}$\thanks{E-mail: faber@kph.tuwien.ac.at},
A. N. Ivanov${^{\,a,b}}$, P. Kienle$^{b,c}$, E.  L. Kryshen${^d}$,
M. Pitschmann${^a}$, N. I. Troitskaya$^{e}$} 
\affiliation{${^a}$Atominstitut der
\"Osterreichischen Universit\"aten, Technische Universit\"at Wien,
Wiedner Hauptstrasse 8-10, A-1040 Wien, \"Osterreich}
\affiliation{${^b}$Stefan Meyer Institut f\"ur subatomare Physik
\"Osterreichische Akademie der Wissenschaften, Boltzmanngasse 3,
A-1090, Wien, \"Osterreich}\affiliation{${^c}$Physik Department, Technische
Universit\"at M\"unchen, D--85748 Garching, Germany} 
\affiliation{${^c}$Petersburg Nuclear
Physics Institute, 188300 Gatchina, Orlova roscha 1, Russian
Federation}\affiliation{ $^d$ Petersburg Nuclear
      Physics Institute, 188300 Gatchina, Orlova roscha 1, Russian
      Federation,\\ $^e$State Polytechnic University of
      St. Petersburg, Polytechnicheskaya 29, 195251, Russian}
\email{ivanov@kph.tuwien.ac.at}

\date{\today}

\begin{abstract}
We calculate the decay rates $\lambda_{\beta^-_c}$ and
 $\lambda_{\beta^-_b}$ of the continuum- and bound-state
 $\beta^-$--decays for bare ${^{205}}{\rm Hg}^{80+}$ and ${^{207}}{\rm
 Tl}^{81+}$ ions. For the ratio of the decay rates $R_{b/c} =
 \lambda_{\beta^-_b}/\lambda_{\beta^-_c}$ we obtain the values
 $R_{b/c} = 0.161$ and $R_{b/c} = 0.190$ for bare ${^{205}}{\rm
 Hg}^{80+}$ and ${^{207}}{\rm Tl}^{81+}$ ions, respectively. The
 theoretical value of the ratio $R_{b/c} = 0.190$ for the decays of
 ${^{207}}{\rm Tl}^{81+}$ agrees within $1\,\%$ of accuracy with the
 experimental data $R^{\exp}_{b/c} = 0.188(18)$, obtained at GSI. The
 theoretical ratio $R_{b/c}= 0.161$ for ${^{205}}{\rm Hg}^{80+}$ is
 about $20\,\%$ smaller than the experimental value $R^{\exp}_{b/c} =
 0.20(2)$, measured recently at GSI.\\ PACS: 12.15.Ff, 13.15.+g,
 23.40.Bw, 26.65.+t
\end{abstract}

\maketitle

As has been shown in \cite{Ivanov1}, the standard theory of weak
interactions of heavy ions \cite{STW}, allows to describe well the
K--shell electron capture ($EC$) and $\beta^+$ decays of the H--like
and He--like ions ${^{140}}{\rm Pr}^{58+}$ and ${^{140}}{\rm
Pr}^{57+}$, agreeing with the experimental data, obtained at GSI
\cite{GSI1}, with an accuracy better than $3\,\%$.

In this letter we apply the standard theory of weak interactions of
heavy ions \cite{STW} and technique developed in \cite{Ivanov1} to the
calculation of the decay rates $\lambda_{\beta^-_c}$ and
$\lambda_{\beta^-_b}$ of the decays 
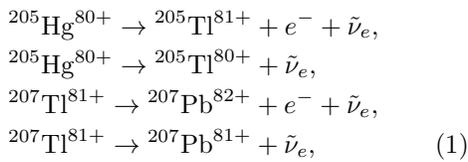
\begin{eqnarray}\label{label1}
 &&{^{205}}{\rm Hg}^{80+} \to
{^{205}}{\rm Tl}^{81+} + e^- + \tilde{\nu}_e,\nonumber\\
&&{^{205}}{\rm
Hg}^{80+} \to {^{205}}{\rm Tl}^{80+} + \tilde{\nu}_e,\nonumber\\
 &&{^{207}}{\rm Tl}^{81+} \to
{^{207}}{\rm Pb}^{82+} + e^- + \tilde{\nu}_e,\nonumber\\
&&{^{207}}{\rm
Tl}^{81+} \to {^{207}}{\rm Pb}^{81+} + \tilde{\nu}_e,
\end{eqnarray}
where ${^{205}}{\rm Hg}^{80+},{^{207}}{\rm Pb}^{82+}$ and
${^{205}}{\rm Tl}^{81+}, {^{207}}{\rm Tl}^{81+}$ are two pairs of bare
ions with quantum numbers $I^P = \frac{1}{2}^-$ and $I^P =
\frac{1}{2}^+$, respectively, and ${^{205}}{\rm Tl}^{80+}$ and
${^{207}}{\rm Pb}^{81+}$ are the H--like ions. The continuum- and
bound-state $\beta^-$--decays in Eq.(\ref{label1}) satisfy the
selection rule $\Delta I^P = 0^-$, which corresponds to the selection
rule of the first--forbidden $\beta$--decays \cite{EK66}.  The
bound-state $\beta^-$--decay to the K--shell is the time reversed
orbital K--shell Electron Capture ($EC$) decay, which we analysed in
\cite{Ivanov1}. The main distinction of the bound state
$\beta^-$--decay from the $EC$--decay is that the bound electron can
be not only on the K--shell in the $1s$ state but on any other shells
in any excited $ns$ state, the contribution of which is about
$20\,\%$.

A measurement of the continuum- and bound-state $\beta^-$ decays of
 bare ${^{207}}{\rm Tl}^{81+}$ ion was reported by Ohtsubo {\it et
 al.}  \cite{GSI3}.  The experimental value of the ratio of the decay
 rates $R^{\exp}_{b/c} = 0.188(18)$ agrees within one standard
 deviation with the theoretical value $R^{\rm th}_{b/c} = 0.171(1)$
 \cite{FFD}, obtained from the theory employing spectra of allowed
 transitions \cite{GSI3}. Very recently an experimental value
 $R^{\exp}_{b/c} = 0.20(2)$ of the ratio of the continuum- and
 bound-state $\beta^-$ decays of bare ${^{205}}{\rm Hg}^{80+}$ has
 become known \cite{GSI4}. This has motivated us to carry out a
 model--independent calculation of the first--forbidden continuum- and
 bound-state $\beta^-$ decays of bare ${^{205}}{\rm Hg}^{80+}$ and
 ${^{207}}{\rm Tl}^{81+}$ ions.

For the calculation of the rates of the $\beta^-$--decays we use the
Hamilton density operator \cite{Ivanov1,STW}
\begin{eqnarray}\label{label2}
 {\cal H}_W(x) &=& \frac{G_F}{\sqrt{2}}\,V_{ud}\,
  [\bar{\psi}_p(x)\gamma^{\mu}(1 - g_A\gamma^5)\,\psi_n(x)]\nonumber\\
  &&\times\,[\bar{\psi}_e(x) \gamma_{\mu}(1 - \gamma^5)\psi_{\nu_e}(x)]
  + {\rm h.c.}
\end{eqnarray}
with standard notations \cite{Ivanov1,STW,PDG06}. In our calculations
the anti--neutrino is assumed to be a massless Dirac anti--particle as
in \cite{Ivanov1}.
\subsection*{The continuum-state  $\beta^-$--decay}
In the rest frame of the mother ion the amplitudes of the
continuum-state $\beta^-$--decay are defined by
\begin{eqnarray}\label{label3}
  \hspace{-0.3in}&& M_{II_z \to I'I'_z} = \nonumber\\
\hspace{-0.3in}&& = - {_{I'I'_z}}\langle
   \tilde{\nu}_e(\vec{k}\,)e^-(\vec{p}_-) d(\vec{q}\,)|{\cal
   H}_W(0)|m(\vec{0}\,)\rangle_{II_z},
\end{eqnarray}
where $II_z = \frac{1}{2},\pm \frac{1}{2}$ and $I'I'_z =
\frac{1}{2},\pm \frac{1}{2}$ determine the spinorial states of the
mother $m$ and daughter $d$ ions. The decay rate
$\lambda_{\beta^-_c}$ of the continuum-state $\beta^-$-decay is
defined by \cite{Ivanov1}
\begin{eqnarray}\label{label4}
 \hspace{-0.3in}&&\lambda_{\beta^-_c} = \frac{1}{2
M_m}\int\frac{d^3\vec{q}}{(2\pi)^3 2 E_d}\frac{d^3\vec{p}_-}{(2\pi)^3
2 E_-}\frac{d^3\vec{k}}{(2\pi)^3 2 E_{\tilde{\nu}_e}} \nonumber\\
 \hspace{-0.3in}&&\times\,(2\pi)^4\delta^{(4)}(k + p_- + q - k_m)\,
\,F(Z + 1, E_-)\nonumber\\
\hspace{-0.3in}&&\times\,\frac{1}{2}\sum_{I_z, I'_z, \sigma_-}
|M_{I I_z \to I' I'_z}|^2,
\end{eqnarray}
where $k = (E_{\tilde{\nu}_e}, \vec{k}\,)$, $p_- = (E_-, \vec{p}_-)$,
$q = (E_d, \vec{q}\,)$ and $k_m = (M_m,\vec{0}\,)$ are 4--momenta of
the interacting particles, $F(Z + 1, E_-)$ is the Fermi function
\cite{Ivanov5}
\begin{eqnarray}\label{label5}
\hspace{-0.3in}&&F(Z + 1, E_-) = \Big(1 +
\frac{1}{2}\gamma\Big)\,\frac{4(2 R p_-)^{2\gamma}}{\Gamma^2(3 +
2\gamma)}\nonumber\\
\hspace{-0.3in}&&\times\,e^{\,\textstyle +\,\frac{\pi (Z + 1)\alpha
E_-}{p_-}}\Big|\Gamma\Big(1 + \gamma + i\,\frac{\alpha (Z + 1)
E_-}{p_-}\Big)\Big|^2,
\end{eqnarray}
with $\gamma = \sqrt{1 - ((Z + 1) \alpha)^2} - 1$, $p_- = \sqrt{E^2_-
  - m^2_e}$, $Z = 80$ and $Z = 81$ for ${^{205}}{\rm Hg}^{80+}$ and
${^{207}}{\rm Tl}^{81+}$, respectively. The summation in
Eq.(\ref{label4}) should be carried out over all polarisations of the
interacting particles, where $\sigma_- = \pm \frac{1}{2}$ is a
polarisation of the electron. The anti--neutrino is polarised parallel
to the momentum $\vec{k}$.  Following \cite{Ivanov1}, for the
amplitudes of the continuum-state $\beta^-$--decay we get the
expressions
\begin{eqnarray}\label{label6}
\hspace{-0.3in}&&M_{\frac{1}{2},+\frac{1}{2}\to
  \frac{1}{2},+\frac{1}{2}}  = \sqrt{2
  M_m 2 E_d}\,{\cal M}_{m \to d} \nonumber\\
  \hspace{-0.3in}&&\times\,
 [\bar{u}_e(\vec{p}_-,\sigma_-)\,(g_A\gamma^0 - \gamma^3)\,(1 - \gamma^5)
  v_{\tilde{\nu}_e}(\vec{k},+\frac{1}{2})],\nonumber\\
\hspace{-0.3in}&&M_{\frac{1}{2},+\frac{1}{2}\to
  \frac{1}{2},-\frac{1}{2}} =
\sqrt{2 M_m 2 E_d}\,{\cal M}_{m \to d} \nonumber\\
  \hspace{-0.3in}&&\times\,
 [\bar{u}_e(\vec{p}_-,\sigma_-)\,(\gamma^1 + i \gamma^2)\,(1 - \gamma^5)
  v_{\tilde{\nu}_e}(\vec{k},+\frac{1}{2})],\nonumber\\
\hspace{-0.3in}&&M_{\frac{1}{2},-\frac{1}{2}\to
  \frac{1}{2},+\frac{1}{2}} = \sqrt{2 M_m 2
  E_d}\,{\cal M}_{m \to d} \nonumber\\
  \hspace{-0.3in}&&\times\,
 [\bar{u}_e(\vec{p}_-,\sigma_-)\,(\gamma^1 -  i\gamma^2)\,(1 - \gamma^5)
  v_{\tilde{\nu}_e}(\vec{k},+\frac{1}{2})],\nonumber\\
\hspace{-0.3in}&&M_{\frac{1}{2},-\frac{1}{2}\to
  \frac{1}{2},-\frac{1}{2}} =  \sqrt{2 M_m 2
  E_d}\,{\cal M}_{m \to d} \nonumber\\
  \hspace{-0.3in}&&\times\,
 [\bar{u}_e(\vec{p}_-,\sigma_-)\,(g_A\gamma^0 +  \gamma^3)\,(1 - \gamma^5)
  v_{\tilde{\nu}_e}(\vec{k},+\frac{1}{2})].
\end{eqnarray}
where $\bar{u}_e$ and $v_{\tilde{\nu}_e}$ are Dirac bispinors of the
electron and the anti--neutrino, ${\cal M}_{m\to d}$ is the nuclear
matrix element defined by
\begin{eqnarray}\label{label7}
{\cal M}_{m\to d} = -\, \frac{G_F}{\sqrt{2}}\,V_{ud}\int
  d^3x\,\Psi^*_d(\vec{r}\,)\,\Psi_m(\vec{r}\,),
\end{eqnarray}
where $\Psi_d(\vec{r}\,)$ and $\Psi_m(\vec{r}\,)$ are the wave
functions of the daughter and mother nuclei. For the numerical
calculations we assume that the product
$\Psi^*_d(\vec{r}\,)\,\Psi_m(\vec{r}\,)$ has the Wood--Saxon shape
\cite{Ivanov1}. Substituting the amplitudes Eq.(\ref{label6}) into
Eq.(\ref{label4}) and carrying out the summation over polarisations we get
\begin{eqnarray}\label{label8}
\lambda_{\beta^-_c} = (3 + g^2_A)\,\frac{|{\cal M}_{m \to d}|^2}{\pi^3}
 f(Q_{\beta^-_c}, Z + 1),
\end{eqnarray}
where the Fermi integral $f(Q_{\beta^-_c}, Z + 1)$ is 
\begin{eqnarray}\label{label9} 
\hspace{-0.3in}&& f(Q_{\beta^-_c}, Z + 1) = \int^{Q_{\beta^-_c} +
m_e}_{m_e}(Q_{\beta^-_c} + m_e - E_-)^2\nonumber\\
\hspace{-0.3in}&&\times\,F(Z + 1,E_-)\,\sqrt{E^2_- - m^2_e}\,E_- dE_-
\end{eqnarray}
The $Q$--values of the continuum-state $\beta^-$--decays are equal to
$Q_{\beta^-_c}= 1515.734\,{\rm keV}$ and $Q_{\beta^-_c} =
1407.471\,{\rm keV}$ for ${^{205}}{\rm Hg}^{80+}$ and ${^{207}}{\rm
Tl}^{81+}$, respectively \cite{Ivanov5,Ivanov5a,NDC}.
\subsection*{The bound-state  $\beta^-$--decay}
In the bound-state $\beta^-$--decay the electron in the final state
can be in any bound $ns$--state. Due to hyperfine interactions the
$ns$--state is splitted into two hyperfine states $(ns)_{F = 0}$ and
$(ns)_{F = 1}$ with a total spin $F = 0$ and $F = 1$, respectively
\cite{HFS1}.

The experimental value of the hyperfine energy level splitting for the
$1s$--state $\Delta E^{\exp}_{1s} = E_{(1s)_{F = 0}} - E_{(1s)_{F =
1}} = -\,3.24409 \pm 0.00029\,{\rm eV}$ \cite{HFS2} agrees well with
the theoretical one $\Delta E^{\rm th}_{1s} = - 3.275\,{\rm eV}$
\cite{HFS1}. The hyperfine splitting $\Delta E^{\rm th}_{ns}$ of the
energy level of the excited $ns$ state is related to $\Delta E^{\rm
th}_{1s}$ as \cite{HFS1}
\begin{eqnarray}\label{label10}
\hspace{-0.3in}\Delta E_{ns} &=& \frac{\Delta E_{1s}}{(3 + 2\gamma)[(n
+ \gamma)^2 - \gamma(2 + \gamma)]^2}\nonumber\\
\hspace{-0.3in}&\times&\Big[2(n + \gamma) + \sqrt{(n +
\gamma)^2 - \gamma(2 + \gamma)}\Big].
\end{eqnarray}
The decay rate $\lambda_{\beta^-_b}$ of the bound-state
$\beta^-$--decay into any hyperfine states $(ns)_{F = 0}$ and $(ns)_{F
= 1}$ is defined by
\begin{eqnarray}\label{label11}
\hspace{-0.3in}&&\lambda_{\beta^-_b} = \sum^{\infty}_{n = 1}\sum_{F =
0,1}\lambda_{\beta^-_b}((ns)_F) = \frac{1}{2 M_m}\nonumber\\
\hspace{-0.3in}&&\times\int \frac{d^3 \vec{q}}{(2\pi)^3 2
E_d}\frac{d^3\vec{k}}{(2\pi)^3 2 E_{\tilde{\nu}_e}}\,(2\pi)^4
\delta^{(4)}(q + k - k_m) \nonumber\\
\hspace{-0.3in}&&\times \,\frac{1}{2}\sum^{\infty}_{n = 1}\sum_{F = 0,
1}\sum_{I_z, M_F}|M^{(n)}_{I I_z \to FM_F}|^2,
\end{eqnarray}
where $k = (E_{\tilde{\nu}_e}, \vec{k}\,)$, $q = (E_d, \vec{q}\,)$ and
$k_m = (M_m,\vec{0}\,)$ are 4--momenta of the interacting
particles. In the rest frame of the mother ion the amplitudes of the
bound-state $\beta^-$--decay are determined by
\begin{eqnarray}\label{label12}
  \hspace{-0.2in}&&M^{(n)}_{II_z \to F M_F} = - {_{F M_F}}\langle
   \tilde{\nu}_e(\vec{k}\,)d^{(n)}(\vec{q}\,)|{\cal
     H}_W(0)|(\vec{0}\,)\rangle_{II_z},\nonumber\\
\hspace{-0.2in}&&
\end{eqnarray}
Following \cite{Ivanov1} for the amplitudes of the transitions $II_z
\to FM_F$ we get the expressions
\begin{eqnarray}\label{label13}
\hspace{-0.3in}&&M^{(n)}_{\frac{1}{2}, - \frac{1}{2}\to 0,0} = \sqrt{2 M_m
  2 E_d E_{\tilde{\nu}_e}}\, (3 -
  g_A) \nonumber\\
\hspace{-0.3in}&&\times\,\sqrt{2}\,{\cal M}_{m \to d}\, \langle
  \psi^{(Z+1)}_{ns}\rangle,\nonumber\\
\hspace{-0.3in}&&M^{(n)}_{\frac{1}{2},+\frac{1}{2}\to 1, +1} = \sqrt{2 M_m 2
 E_d E_{\tilde{\nu}_e}}\,(1 + g_A) \nonumber\\
\hspace{-0.3in}&&\times\, 2\,{\cal M}_{m \to d}\,\langle
 \psi^{(Z+1)}_{ns}\rangle,\nonumber\\
\hspace{-0.3in}&&M^{(n)}_{\frac{1}{2},- \frac{1}{2}\to 1,0} = \sqrt{2 M_m 2 E_d(\vec{q}\,)
E_{\tilde{\nu}_e}(\vec{k}\,)}\,(1 +
g_A)\nonumber\\
\hspace{-0.3in}&&\times\,\sqrt{2}\,{\cal M}_{m \to d}\,\langle
\psi^{(Z + 1)}_{ns}\rangle,
\end{eqnarray}
where $\langle \psi^{(Z + 1)}_{ns}\rangle$ is the wave function of the
bound electron in the $ns$--state, averaged over the nuclear density
\cite{Ivanov1}. The wave function $\psi^{(Z + 1)}_{ns}$ is the
solution of the Dirac equation \cite{RWF1,RWF2}. Substituting the
amplitudes Eq.(\ref{label13}) into Eq.(\ref{label11}) we get the decay
rate
\begin{eqnarray}\label{label14}
\hspace{-0.3in}&&\lambda_{\beta^-_b} = (3 - g_A)^2 |{\cal M}_{m \to
d}|^2\sum^{\infty}_{n = 1} |\langle
\psi^{(Z+1)}_{ns}\rangle|^2\,\frac{Q^2_{ns}}{2\pi}\nonumber\\
\hspace{-0.3in}&& + 3(1 + g_A)^2 |{\cal M}_{m \to
d}|^2\sum^{\infty}_{n = 1} |\langle
\psi^{(Z+1)}_{ns}\rangle|^2\,\frac{Q^2_{ns}}{2\pi}\nonumber\\
\hspace{-0.3in}&& = 2 (3 + g^2_A)\,|{\cal M}_{m \to d}|^2
\sum^{\infty}_{n = 1}|\langle
\psi^{(Z+1)}_{ns}\rangle|^2\,\frac{Q^2_{ns}}{\pi},
\end{eqnarray}
where $Q_{ns}$ is the $Q$--value of the bound-state $\beta^-$--decays
into the bound $ns$--state \cite{Ivanov5b}.
\subsection*{The ratio of the continuum- and  bound-state  $\beta^-$--decays}
The ratio of the continuum- and bound-state $\beta^-$--decay rates is
\begin{eqnarray}\label{label15}
\hspace{-0.3in}R_{\beta^-_b/\beta^-_c} = \sum^{\infty}_{n = 1}\frac{2\pi^2 Q^2_{\rm
ns}|\langle \psi^{(Z+1)}_{ns}\rangle|^2}{f(Q_{\beta^-_c}, Z + 1)}.
\end{eqnarray}
The numerical values of the ratio of the $\beta^-$--decays of
${^{205}}{\rm Hg}^{80+}$ and ${^{207}}{\rm Tl}^{81+}$ are 
\begin{eqnarray}\label{label16}
\hspace{-0.3in}{^{205}}{\rm Hg}^{80+}:&&R^{\rm th}_{\beta^-_b/\beta^-_c} =
0.161,\nonumber\\
\hspace{-0.3in}{^{207}}{\rm Tl}^{81+}:&&R^{\rm th}_{\beta^-_b/\beta^-_c} = 0.190.
\end{eqnarray}
These results, obtained for $R = 1.1\times A^{1/3}\,{\rm fm}$
\cite{Ivanov1,HFS2}, are stable under reasonable variations of $R$ as
it has been observed for the $EC$ and $\beta^+$ decays of the H--like
${^{140}}{\rm Pr}^{58+}$ and He--like ${^{140}}{\rm Pr}^{57+}$ ions in
\cite{Ivanov1}.
\subsection*{Concluding discussion}
Using the standard theory of weak interactions of heavy ions we have
calculated the decay rates of the continuum- and bound-state
$\beta^-$--decays of bare ${^{205}}{\rm Hg}^{80+}$ and ${^{207}}{\rm
Tl}^{81+}$ ions. These are first --forbidden $\beta$--decays
\cite{EK66}. Our result for the ratio of the $\beta^-$--decays of bare
${^{207}}{\rm Tl}^{81+}$ ion $R^{\rm th}_{\beta^-_b/\beta^-_c} =
0.190$ agrees with the experimental value
$R^{\exp}_{\beta^-_b/\beta^-_c} = 0.188(18)$ within an accuracy of
about $1\,\%$. This is much better than the theoretical value $R^{\rm
th}_{\beta^-_b/\beta^-_c} = 0.171(1)$, obtained by Takahashi and Yokoi
\cite{FFD}.

Our result for the ratio of the $\beta^-$--decay rates of bare
${^{205}}{\rm Hg}^{80+}$ ions deviates from the experimental value
$R^{\exp}_{\beta^-_b/\beta^-_c} = 0.20(2)$ by about $20\,\%$. Since
our agreement with the experimental value of
$R^{\exp}_{\beta^-_b/\beta^-_c} = 0.188(18)$ for bare ${^{207}}{\rm
Tl}^{81+}$ ions is about of $1\,\%$, we can argue that the
experimental value for the ratio of the $\beta^-$--decay rates of bare
${^{205}}{\rm Hg}^{80+}$ ions maybe actually too high. Our assertion
is based on the following. The dependence of the ratio
$R_{\beta^-_b/\beta^-_c}$ on the electric charge is rather
weak. Indeed, for $Z = 81$ instead of $Z = 80$ the ratio of the
$\beta^-$--decays $R_{\beta^-_b/\beta^-_c}$ of bare ${^{205}}{\rm
Hg}^{80+}$ ions changes to the value $R_{\beta^-_b/\beta^-_c} =
0.167$. Thus, the main distinction is due to different $Q$--values of
the continuum- and bound-state $\beta^-$--decays.  The $Q$--values
$Q_{\beta^-_c}= 1515.734\,{\rm keV}$ and $Q_{\beta^-_c} =
1407.471\,{\rm keV}$ of continuum-state $\beta^-$--decays result in
the Fermi integrals $f(Q_{\beta^-_c}, Z + 1) = 22.119\,{\rm MeV}^5$
and $f(Q_{\beta^-_c}, Z + 1) = 17.747\,{\rm MeV}^5$ for ${^{205}}{\rm
Hg}^{80+}$ and ${^{207}}{\rm Tl}^{81+}$, respectively. On the other
hand the bound-state $\beta^-$--decay rates scale with $Q^2_{ns}$,
where $Q_{1s} = 1614.557\,{\rm keV}$ and $Q_{1s} = 1509.053\,{\rm
keV}$ for ${^{205}}{\rm Hg}^{80+}$ and ${^{207}}{\rm Tl}^{81+}$,
respectively. This implies that the ratio of the continuum- and
bound-state $\beta^-$--decay rates for bare ${^{205}}{\rm Hg}^{80+}$
ions should be as minimum $1.1$ times smaller relative to the ratio of
the continuum- and bound-state $\beta^-$--decay rates for bare
${^{207}}{\rm Tl}^{81+}$ ions. This gives $R_{\beta^-_b/\beta^-_c}
\sim 0.171$ for bare ${^{205}}{\rm Hg}^{80+}$ ions at the nuclear
radius $R = 1.1\times A^{1/3}$. A reduction of the value
$R_{\beta^-_b/\beta^-_c} \sim 0.171$ to $R_{\beta^-_b/\beta^-_c}=
0.161$ is caused by the effective densities of electrons in the
$ns$--states for different $(Z + 1)$ values of electric charges of the
ions ${^{205}}{\rm Tl}^{80+}$ and ${^{207}}{\rm Pb}^{81+}$ in the
final state of the bound-state $\beta^-$--decays.


\begin{thebibliography}{9}
\bibitem{Ivanov5a} The $Q$--value of the continuum--state
$\beta^-$--decays we calculate as: $Q_{\beta^-_c} = (\Delta_m
- \Delta_d) + (\epsilon^{(Z + 1)}_{TF} - \epsilon^{(Z)}_{TF}) =
(-22304 + 23835) + (- 521.826 + 506.560) \,{\rm keV} = 1515.734\,{\rm
keV}$ for ${^{205}}{\rm Hg}^{80+} \to {^{205}}{\rm Tl}^{81+} + e^- +
\tilde{\nu}_e$ and $Q_{\beta^-_c} = (\Delta_m - \Delta_d) +
(\epsilon^{(Z + 1)}_{TF} - \epsilon^{(Z)}_{TF}) = (-21044 + 22467) +
(- 537.355 + 521.826) \,{\rm keV} = 1407.471\,{\rm keV}$ for
${^{207}}{\rm Tl}^{81+} \to {^{207}}{\rm Pb}^{82+} + + e^- +
\tilde{\nu}_e$, where $\epsilon^{(Z)}_{\rm TF} =
-\,14.33\,Z^{2.39}\,{\rm eV}$ is the binding energy of electrons in
the atom with $Z$ electrons, defined in the Thomas--Fermi model (see
also \cite{NDC}).
\bibitem{Ivanov5b} The $Q$--value of the bound-state $\beta^-$--decay
is defined by $Q_{ns} = Q_{\beta^-_c} + m_e(1 - \frac{n +
\gamma}{\sqrt{n^2 + 2 (n-1)\gamma}})$.
\end{thebibliography}
\end{document}